\documentclass[a4paper,11pt]{article}
\pdfoutput=1 % if your are submitting a pdflatex (i.e. if you have
\usepackage{jcappub} % for details on the use of the package, please
\usepackage[T1]{fontenc} % if needed

\newcommand{\Hconf}{\mathcal{H}}
\newcommand{\beq}{\begin{equation}}
\newcommand{\eeq}{\end{equation}}

\newcommand{\htwo}{\hspace{2pt}}
\newcommand{\nhat}{\hat{\textbf{n}}}
\newcommand{\del}{\partial}

\newcommand{\WMAP}{{\slshape WMAP~}}
\newcommand{\LCDM}{$ \Lambda $CDM~}

\newcommand{\Planck}{{\slshape Planck~}}
\newcommand{\Planckc}{{\slshape Planck}}
\usepackage{tensor}
\usepackage{todonotes}
\usepackage{enumerate}
\usepackage{xspace}
\usepackage{arydshln}
\usepackage{subfigure}

\title{Constraining dark sector perturbations II: ISW and CMB lensing tomography}

\author[a,b]{B. Soergel,}
\author[a,b,c]{T. Giannantonio,}
\author[b,d,e]{J. Weller}
\author[f]{and R. A. Battye}

\affiliation[a]{Kavli Institute for Cosmology and Institute of Astronomy, University of Cambridge, Madingley Road, Cambridge {CB3 0HA}, United Kingdom}
\affiliation[b]{University Observatory, Department of Physics, Ludwig-Maximilians-University Munich,  \\Scheinerstr. 1, D-81679 M\"unchen, Germany}
\affiliation[c]{Centre for Theoretical Cosmology, DAMTP, University of Cambridge, Wilberforce Road, Cambridge CB3 0WA, United Kingdom}
\affiliation[d]{Excellence Cluster Universe, Boltzmannstr. 2, D-85748 Garching, Germany}
\affiliation[e]{Max Planck Institute for Extraterrestrial Physics,\\ Giessenbachstr. 1, D-85748 Garching, Germany}
\affiliation[f]{Jodrell Bank Centre for Astrophysics, School of Physics and Astronomy, The University of Manchester, Manchester M13 9PL, United Kingdom}

\emailAdd{bsoergel@ast.cam.ac.uk}
\emailAdd{t.giannantonio@ast.cam.ac.uk}
\emailAdd{jochen.weller@usm.lmu.de}
\emailAdd{richard.battye@manchester.ac.uk}

\abstract{
Any Dark Energy (DE) or Modified Gravity (MG) model that deviates from a cosmological constant requires a consistent treatment of its perturbations, which
can be described in terms of an effective entropy perturbation and an anisotropic stress.
We have considered a recently proposed generic parameterisation of DE/MG perturbations 
and compared it to data from the \Planck satellite and six galaxy catalogues, including temperature-galaxy ($Tg$), CMB lensing-galaxy ($\varphi g$) and galaxy-galaxy ($gg$) correlations.
Combining these observables of structure formation with tests of the background expansion allows us to investigate the properties of DE/MG both at the background and the perturbative level.
Our constraints on DE/MG are mostly in agreement with the cosmological constant paradigm, while we also find that the constraint on the equation of state $w$ (assumed to be constant) depends on the model assumed for the perturbation evolution. 
We obtain $w=-0.92^{+0.20}_{-0.16}$ ($95\%$ CL; CMB+$gg$+$Tg$) in the entropy perturbation scenario; in the anisotropic stress case the result is $w=-0.86^{+0.17}_{-0.16}$. Including the lensing correlations shifts the results towards higher values of $w$. 
If we include a prior on the expansion history from recent Baryon Acoustic Oscillations (BAO) measurements, we find that the constraints tighten closely around $w=-1$, making it impossible to measure any DE/MG perturbation evolution parameters. If, however, upcoming observations from surveys like DES, Euclid or LSST show indications for a deviation from a cosmological constant, our formalism will be a useful tool towards model selection in the dark sector.
}

\begin{document}
\maketitle
\flushbottom

\section{Introduction}
The discovery of the accelerating expansion of the Universe \cite{Perlmutter1999, Riess1998} has led to the development of a host of theories designed to explain it. This can be done either by adding an additional component to the energy content of the Universe or alternatively by modifying the equations of gravity on cosmological scales. The former is known as \emph{Dark Energy} (DE), the latter is commonly referred to as \emph{Modified Gravity} (MG). 
Both of these main classes contain a vast number of different models, see e.g. \cite{Amendola2010} and \cite{Clifton2012} for an overview of individual theories.
At the background level, both DE and MG can be mapped onto a fluid with an effective equation of state $P = w(a) \rho$. Hence, using only measurements of the expansion of the cosmological background, DE and MG remain indistinguishable.  However, the situation improves at the perturbative level. Better discrimination between DE and MG could be possible at the stage of linear perturbations, so their evolution is a useful probe for the origin of the cosmic acceleration. It has been shown that in order to consistently calculate the perturbation evolution in general DE/MG models, one must also take into account perturbations in this component, see e.g.\@\xspace \cite{Weller2003} and \cite{Bean2003}. Given the plethora of different models, it is crucial to find a model-independent treatment not only at the background level, but also for the evolution of perturbations; in recent years, this field has been extensively studied theoretically and observationally, see e.g. 
 \cite{Linder2005,Hu2007,Bertschinger2008,Zhao2008,Hu2008,Fang2008,Giannantonio2009,Skordis2009,Bean2010,Baker2011,Hojjati2011,Amendola2012,Baker2012,Battye2012,Sawicki2012,Dossett2013,Motta2013,Battye2013a,Battye2013b,Hu2013,Frusciante2013,Bloomfield2013,Amendola2013,Cardona2014}.
The DE/MG parameterisation recently developed in \cite{Battye2012, Battye2013a, Battye2013b} is such a model-independent approach, which has the added value of being based on a field theory with only few generic assumptions, so that it does not require any heuristic modifications of the perturbative evolution equations.

Currently the tightest constraints on DE/MG are obtained by breaking the geometric degeneracy in the \emph{Planck}  \cite{Planck2013} or \emph{WMAP} \cite{Hinshaw2013} cosmic microwave background (CMB) measurements by adding supernovae Ia (SNe Ia) or baryon acoustic oscillations (BAO) data. Nevertheless, these background probes only allow us to measure $w(a)$, but not to test the perturbative effects of DE/MG, such as structure formation.
 By contrast, the late-time {integrated Sachs-Wolfe} (ISW) {effect} \cite{Sachs1966} in the large-scale CMB temperature anisotropies is sensitive to the properties of DE/MG beyond the simple equation of state: while the CMB photons are propagating through an overdensity, its gravitational potential gets shallower due to the accelerated expansion, which leads to a net temperature gain, and vice versa.
This signature is sensitive to the evolution of the potentials and therefore to the perturbative properties of the model of cosmic acceleration.
As DE/MG only becomes important for the perturbation evolution at late cosmological times, the ISW effect is only present on the largest angular scales and thus its constraining power is limited by cosmic variance. Cross-correlating CMB maps with large-scale structure (LSS) data however enhances the detection significance as the galaxy distribution also traces the gravitational potential \cite{Crittenden1995}. This attenuates the limitations due to cosmic variance, and also makes the ISW effect a useful probe for constraining the redshift evolution of DE and MG \cite{Ho2008, Giannantonio2008,Giannantonio2009, Giannantonio2012, Giannantonio2013, Planck2013a, Munshi2014}.

Another secondary CMB effect sensitive to DE/MG at the perturbative level is the weak gravitational lensing that occurs as the CMB photons travel through the LSS \cite{2006PhR...429....1L}; this has been detected and mapped by the Atacama Cosmology Telescope \cite{2011PhRvL.107b1301D}, the South Pole Telescope \cite{2012ApJ...756..142V} and \Planck \cite{PlanckXVII}.
Also in this case, it is useful to cross-correlate the CMB lensing maps with LSS data, in order to study the redshift evolution of the signal \cite{2012PhRvD..86h3006S, 2012ApJ...753L...9B, PlanckXVII, Giannantonio2013b}. 
Finally, the galaxy-galaxy correlation functions, which are equivalent to using the full shape of the galaxy power spectrum, are also sensitive to DE/MG at the perturbative level, whereas BAOs only test the background expansion. These three data sets together form a powerful combination for investigating the origin of the cosmic acceleration.

Current and upcoming projects like the \emph{Dark Energy Survey} (DES) \cite{Abbott2005}, \emph{Euclid} \cite{ESA2011} and the \emph{Large Synoptic Survey Telescope} (LSST) \cite{LSST2012} will provide LSS and weak lensing data of unprecedented quality. It is therefore important to develop a framework to make full use of the upcoming wealth of observations, especially by correlating  and combining different probes, and by using model-independent parameterisations of DE and MG. This can in principle allow to rule out entire classes of theories instead of testing single models.
In this paper, we constrain the DE/MG models by \cite{Battye2012, Battye2013a, Battye2013b} using
the combined measurements by \cite{Giannantonio2012, Giannantonio2013, Giannantonio2013b}, that include LSS, ISW, and CMB lensing correlations with six galaxy catalogues. Our analysis is complementary to the one by Ref. \cite{Battye2014}, in which measurements of cosmic shear and the CMB lensing auto-spectrum are used to constrain the same DE/MG parameterisation.

When not using the BAO data, we are able to rule out a significant part of the DE/MG parameter space, constraining deviations from the standard model, whereas  BAO data limit us to the degenerate case around $w=-1$, thus preventing us from obtaining interesting constraints. Furthermore, we find that the constraint on the background equation of state depends on the model assumed for the perturbation evolution. 
Our work is structured as follows: in Section~\ref{sec:models} we outline the most important features of the parameterisation of DE/MG perturbations.
In Section~\ref{sec:dataset} we describe the combined data set we use, and we present our results in Section~\ref{sec:mcmcresults}, before concluding in Section~\ref{sec:conclusion}.

\section{Theoretical models}
\label{sec:models}
At the background level, all DE and MG models can effectively be described by one function of time, the equation of state $w(a)$. Given that current data is still not constraining this quantity particularly well, the equation of state is commonly assumed to be either a constant or alternatively expanded up to linear order in the scale factor $a$ \cite{Chevallier2001,Linder2003} $w(a) = w_0 + (1-a)w_a $.
The formalism recently developed in \cite{Battye2012, Battye2013a, Battye2013b} aims to provide a similar tool, i.e.\@\xspace a model-independent parameterisation, at the level of linear perturbations. Being a phenomenological approach, it does not require the knowledge of a particular DE or MG Lagrangian, but only imposes some physically motivated restrictions on the full theory space. Therefore this method can be used to constrain a large number of models simultaneously, which is a crucial step towards model selection in the dark sector.

\subsection{Notation and definitions}
We can write the perturbed Einstein equations as
\beq
 \delta_E G_{\mu \nu} = 8 \pi G \left(\delta_E T_{\mu \nu} + \delta_E U_{\mu \nu} \right) \, ,
 \eeq 
where $\delta_E$ is a Eulerian perturbation (see \cite{Battye2012}) and $U_{\mu \nu}$ is the DE/MG energy-momentum-tensor. We decompose its perturbations
\beq
\delta_E U\indices{^\mu_\nu} = \delta \rho \htwo  u^\mu u_\nu + (\rho +P) \htwo \left(v^\mu u_\nu + v_\nu u^\mu \right) + \delta P \htwo \gamma\indices{^\mu_\nu} + P \htwo \tensor{\Pi}{^\mu_\nu}
\eeq
into the standard fluid variables $ \left\{ \delta \rho, v^\mu, \delta P, \Pi\indices{^\mu_\nu}  \right \} $, i.e.\@\xspace into density, velocity and pressure perturbations and the anisotropic stress; the indices $DE/MG$ for the fluid quantities are omitted unless there is a potential ambiguity. Furthermore, $u^\mu$ denotes a time-like unit vector and $\gamma_{\mu\nu}$ is the metric on the 3D spatial hypersurfaces.
We currently restrict ourselves to scalar perturbations (identified by a superscript `s'), because these correspond to density waves that are directly related to observables.
The divergence of the velocity perturbation in Fourier space is $\theta^s \equiv i \textbf{k} \cdot \textbf{v} / k^2 $; this definition differs from the $\theta$ of \cite{Ma1995} by a factor of $-k^2$.
Finally, it is convenient to rewrite $\delta P$ in terms of the effective entropy perturbation
\begin{equation}
w \Gamma = \left (\frac{\delta P}{\delta \rho} -w \right) \delta \, .
\end{equation}
As we are interested in the perturbation evolution, we assume for simplicity $w= \mathrm{const.} = w_0$ for the background expansion;
the formalism is however easily extendable to probe a time-varying $w$.

From the perturbed conservation law $\delta_E (\nabla_\mu U\indices{^\mu_\nu}) = 0$, one obtains the DE/MG fluid equations. In synchronous gauge\footnote{There is a well-known ambiguity in the synchronous gauge that can lead to unphysical gauge modes in the solutions of these equations; see e.g. \cite{Bardeen1980,Press1980,Kodama1984,Mukhanov1990} for a detailed discussion. This issue can be fixed by an appropriate choice of initial conditions \cite{Ma1995}.} they are
\begin{align}
\label{eq:fluid1}
\dot{\delta} &= -(1+w) \left(-k^2 \theta^s + \frac{\dot{h}}{2} \right) - 3 \Hconf w \Gamma \, ;\\
\dot{\theta}^s &= - \Hconf(1-3w)\theta^s - \frac{w}{1+w} \left( \delta + \Gamma  -\frac{2}{3} \Pi^s \right) \, ,
\label{eq:fluid2}
\end{align}
where $h$ and $\eta$ are the metric perturbations and dots denote derivatives with respect to conformal time $\tau$. In order to obtain a closed system of differential equations, it is necessary to express two of the four fluid variables in terms of the other two and the metric variables. As the effective entropy perturbation (from now on we will omit the `effective' for brevity) and the anisotropic stress are by construction gauge-invariant variables (see e.g.\@\xspace \cite{Kodama1984,Mukhanov1990}), they are a sensible choice for this, so we write them as
\beq 
\{\Gamma, \Pi^s\}=\{\Gamma, \Pi^s\}(\delta, \theta^s, h, \eta) \, .
\eeq 
Given that we work only at the level of linear perturbations, this will lead to relations of the type 
\beq
\Gamma = A \delta + B \theta^s + C \dot{h} + ... 
\eeq
 and similarly for $\Pi^s$. Following \cite{Battye2013a, Battye2013b} we will call these the \emph{equations of state} of the DE/MG perturbations,  as they are the analogues of the background equation of state $P=w \rho$.

\subsection{Equations of state}
\label{subsec:eos}
In this section, we briefly describe the two classes of models that we study: the \emph{entropy perturbation} and the \emph{anisotropic stress model}. For a detailed discussion, see Refs. \cite{Battye2012, Battye2013a, Battye2013b}.

\subsubsection{Entropy perturbation model}
In the derivation of the equations of state the following assumptions have been made \cite{Battye2013b}:\linebreak
(1) The field content of the dark sector is 
$\mathcal{L} = \mathcal{L}(\phi,\del_\mu \phi, \del_\mu \del_\nu \phi, g_{\mu\nu}, \del_\alpha g_{\mu\nu})$
and there is at most a linear dependence on $\del_\alpha g_{\mu\nu}$. This class of models includes Kinetic Gravity Braiding~\cite{Deffayet2010} but not the general Horndeski theory \cite{Horndeski1974,Deffayet2011,Kobayashi2011}.
(2) The field equations should be at most second order in the perturbed field variables.
(3) The theory should be reparameterisation invariant under $x^\mu \rightarrow x^\mu + \xi^\mu$, i.e.\@\xspace the fluid variables sourcing the gravitational perturbations should be independent of the $\xi^\mu$.

Requiring $\Gamma$ and $\Pi^s$ to be gauge-invariant quantities then leads to the equations of state $w\Pi^s = 0$ and
\begin{align}
\begin{split}
w \Gamma = (\alpha - w) \left[\delta - 3 \mathcal{H}(1+w)\beta_1 \theta^s - \frac{3 \mathcal{H}(1+w)\beta_2}{2 k^2 -6(\dot{\mathcal{H}}-\mathcal{H}^2)} \dot{h} \right. \\
 + \left. \frac{3 \mathcal{H}(1+w)(1-\beta_1 - \beta_2)}{6 \ddot{\mathcal{H}}+ 6 \Hconf^3 -18\Hconf \dot{\Hconf} +2 k^2 \Hconf} \ddot{h} 
\right] \, .
\end{split} 
\label{eq:wGam}
\end{align}
All the DE/MG physics at the level of linear perturbations is now described by the three functions $\alpha$, $\beta_1$ and $\beta_2$. The three are in the most general case functions of time, while $\beta_1$ can also include a quadratic scale dependence. In this work we consider the simplest possible case of time- and scale-independent modifications, where $\alpha$, $\beta_1$ and  $\beta_2$ are constants.
A given DE/MG model can be mapped onto this parameterisation by explicitly computing these functions as demonstrated in \cite{Battye2013b}. However, the main strength of this approach is that it can also be used to probe theories in a model-independent way, i.e.\@\xspace without assuming a specific Lagrangian. A constraint on these functions can potentially rule out entire classes of models.

The interpretation of $\alpha$ is relatively straightforward, as it can be regarded as the square of a generalised sound speed of the DE/MG fluid. In the case of $\alpha \ll 1$, DE/MG tends to cluster, whereas for $\alpha \sim \mathcal{O}(1)$ perturbations in this component decay more quickly.
As shown in \cite{Babichev2007}, there are no problems with causality in the case of $\alpha > 1$. In agreement with this, we find that the perturbation equations do not become unstable in this regime. However, the perturbations decay very fast if $\alpha \gg 1$, so that their contribution to the CMB spectra becomes negligible. For this reason, we choose $0 \leq \alpha \leq 1$ in this work.
In the special case where $\alpha = \mathrm{const.} \equiv \hat{c}_s^2 $, $\beta_1 =1$ and $\beta_2 =0$, the entropy perturbation $w\Gamma$ reduces to a simpler form corresponding to generalized k-essence (e.g. \cite{ArmendarizPicon2000}) models.
If additionally $\alpha = 1$, we have standard quintessence \cite{Zlatev1998}.

Unfortunately, there is no such easy interpretation for the parameters $\beta_1$ and $\beta_2$, so there is also no obvious `physical' range for them. However `unphysical' behaviour will be ruled out by the data; so we can just treat them as phenomenological parameters governing the DE/MG perturbation evolution. Many existing DE/MG models can be mapped onto this formalism: see \cite{Battye2012, Battye2013a, Battye2013b} for details. 

\subsubsection{Anisotropic stress model}
We also consider a model where reparameterisation invariance is explicitly violated, which leads to non-zero anisotropic stress. A simple example for such a theory is the \emph{Elastic Dark Energy} (EDE) model \cite{Battye2007,Battye2013} with $\mathcal{L} = \mathcal{L}(g_{\mu\nu})$. A theory like this can also be interpreted as a massive gravity model: expanding $\mathcal{L}$ up to second order in the perturbations $\delta g_{\mu\nu}$, the only possible term is
\beq
\mathcal{L}^{(2)} = \mathcal{A}^{\alpha \beta \mu \nu} \delta g_{\alpha \beta} \htwo \delta g_{\mu \nu} \,,
 \eeq
which has the same structure as a mass term for the metric perturbations, i.e.\@\xspace a massive graviton. The \emph{mass matrix} $\mathcal{A}^{\alpha \beta \mu \nu}$ encodes different linearised massive gravity theories \cite{Battye2013}. Such a `metric-only' field content results in the equations of state $w\Gamma =0$ and
\beq
w \Pi ^s = \frac{3}{2}\left(w-c_s^2\right)\left[\delta - 3(1+w)(\eta - \eta_0) \right] \, ,
\label{eq:stress}
\eeq
where $\eta_0$ is the initial condition for the synchronous gauge perturbation $\eta$ (see \cite{Battye2007,Bucher1998,Turok1997}).
This scenario is somewhat different from the entropy perturbation model, therefore we will treat them separately in our  analysis. Intuitively one would expect $w > -1$ for this type of model, but we find that the perturbation equations still have linearly stable solutions for $w<-1$ (see also \cite{SawickiVikman2012}). For this reason, we include that region in our analysis. 

The parameter $c_s^2$ can be regarded as the analogue of the sound speed in an elastic medium. In the fluid picture, $c_s^2$ governs how strongly this component tends to cluster. In order to find a prior range of $c_s^2$, we can use the same argument as previously for $\alpha$ and assume $0 \leq c_s^2 \leq 1$. In the case of $w \rightarrow -1$, this model tends towards a cosmological constant and the value of $c_s^2$ becomes irrelevant. If however $w \rightarrow 0$ and $c_s^2 \rightarrow 0$, the model behaves like cold dark matter; see also \cite{Bucher1998} where this model is discussed as \emph{solid dark matter}.

\subsection{Model phenomenology}
\label{subsec:effectofparameters}
 We have implemented the previously described DE/MG perturbation evolution into the publicly available code \textsc{Camb} \cite{Lewis2000, Howlett2012}. In contrast to previous extensions like \textsc{Camb PPF}~\cite{Hu2007,Hu2008,Fang2008,Fang2008b} and \textsc{MGCamb}~\cite{Hojjati2011,Zhao2008}, but similarly to the recently published EFTCAMB \cite{Hu2013,Raveri2014}, our approach is directly motivated and derived from a field theory. The major advantage of mapping the perturbations in the underlying fields onto the fluid variables as described in \cite{Battye2012, Battye2013a, Battye2013b} is however that the latter are directly sourcing the gravitational perturbations. For this reason, generalising existing numerical codes for the evolution of cosmological perturbations is simplified substantially.

 Here we describe the effects of the DE/MG parameterisation assuming a flat universe with baryon and cold dark matter densities $\Omega_b h^2 = 0.027$ and $\Omega_c h^2 = 0.1123$, respectively, a Hubble parameter $h=0.7$ (not to be confused with the synchronous gauge metric perturbation) and an optical depth due to reionisation $\tau_{\mathrm{opt}} = 0.087$. The spectral index and amplitude of the scalar density perturbations are $n_s = 0.963$ and $A_s = 2.441 \times 10^{-9}$ at a pivot scale of $k_p = 0.002\htwo$Mpc$^{-1}$.
In order to compare the theoretical predictions of different models, we choose values for the background equation of state that differ substantially from $w=-1$; this is to better illustrate the effects of the other parameters. 
The new parameters $\{\alpha, \beta_1, \beta_2\}$ can in principle be scale- and time-dependent, but as a first step we treat them as effective values, i.e.\@\xspace as constants.
Another option would be to constrain the DE/MG parameters in a number of $k$- and $a$-bins in a similar manner as what is done for the \textsc{MGCamb} changes to Poisson's equation and anisotropic stress in e.g.\@\xspace~\cite{2010PhRvD..81j3510Z} or for the time evolution of the background equation of state $w(a)$ in e.g.\@\xspace~\cite{Huterer2000}. However, we choose here 
 to keep the number of free parameters as low as possible when using current data, given their limited constraining power.

\subsubsection{Evolution of the potentials}
\begin{figure}[tbp]
 \centering 
\includegraphics[width=0.8\textwidth]{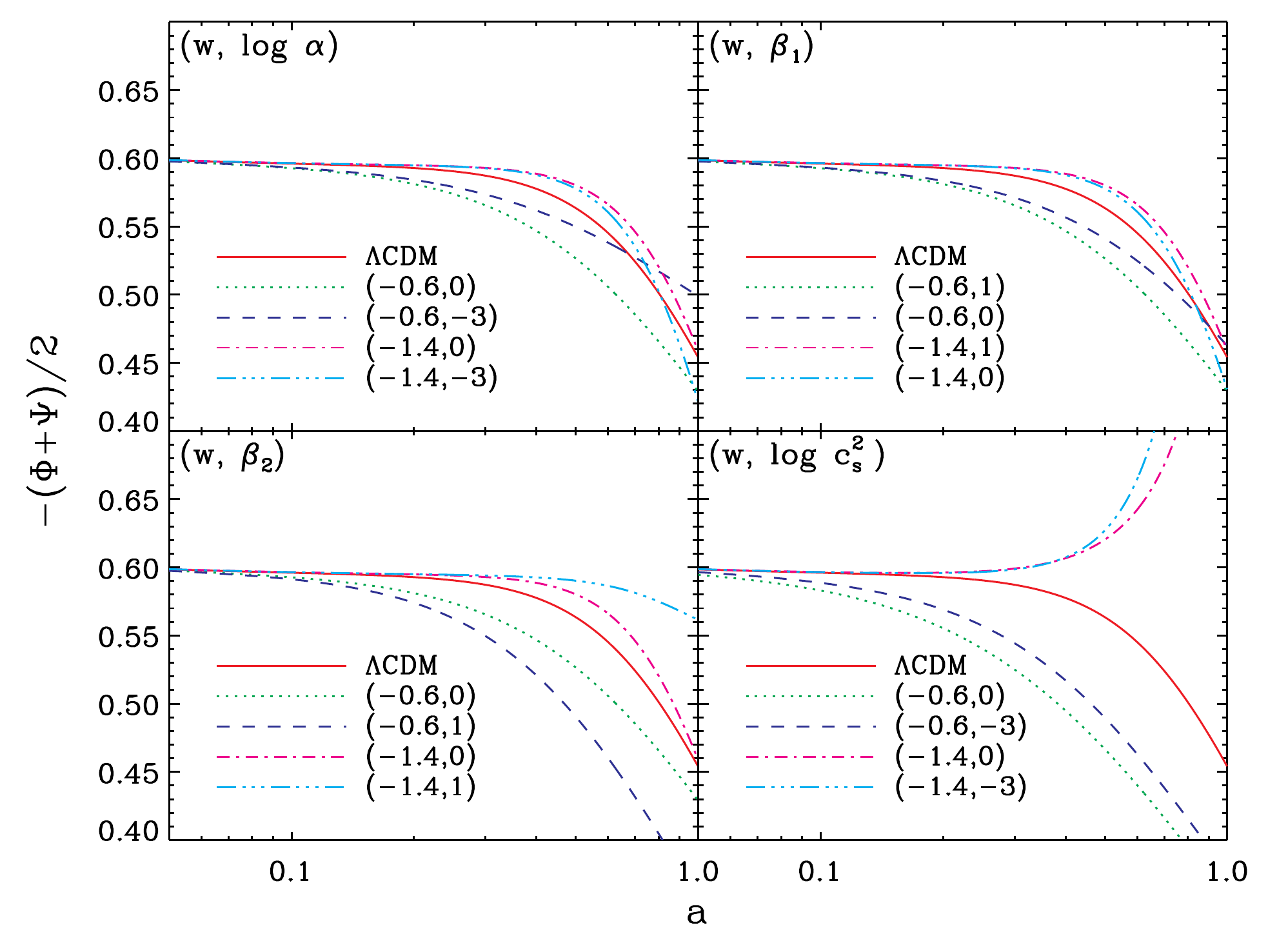}
\caption{\label{fig:Phi} Evolution of the gravitational potentials for the different models we consider: the individual panels show the effect of changing the DE/MG parameters described in the legends. Parameters not explicitly given are set to their fiducial values $\alpha = 1$, $\beta_1 =1$, $\beta_2 = 0$ in the entropy perturbation model and $c_s^2 =1$ in the case with anisotropic stress. The first three panels refer to the former model, while the last panel describes the latter. In all panels, the red line shows the prediction for the fiducial $\Lambda$CDM cosmology.}
\end{figure}
The two Newtonian gauge potentials $\Phi$ and $\Psi$, describing the time and space metric perturbations respectively,
 are related to the synchronous gauge perturbations $h$ and $\eta$ via \cite{Ma1995}
\beq
-\Phi = \frac{1}{2k^2}\left[\ddot{h}+6\ddot{\eta}+\Hconf(\dot{h}+6\dot{\eta})\right]  \, ; \qquad \Psi = \eta - \frac{\Hconf}{2k^2}(\dot{h}+6\dot{\eta}) \, ,
\eeq
which are in turn related to the anisotropic stress via the linearised Einstein equation
\beq
\Phi - \Psi = \frac{8 \pi G \rho a^2}{k^2} w \Pi^s \, .
\eeq
We show in Fig.~\ref{fig:Phi} the evolution of 
the gravitational potentials for one large-scale Fourier mode with $k=0.6\times 10^{-3} \, \text{Mpc}^{-1}$.
In the case without anisotropic stress, i.e.\@\xspace in the first three panels, $(\Phi+\Psi)/2 =\Phi = \Psi$. In the fourth panel, we show the anisotropic stress model.

The effect of $w$ on the evolution of the potentials is straightforward: with a higher (lower) $w$, DE/MG becomes important earlier (later). This translates into an earlier (later) start of the decay of the potential, which is visible in all panels. 
Explaining the impact of the other parameters is more involved. A good way of addressing this is via the linear Poisson equation
\beq
\Phi(k,\tau ) = - \frac{4 \pi G a^2}{k^2}\left[\rho_m(\tau) \htwo \delta_m(k,\tau) + \rho_X(\tau) \htwo \delta_X(k,\tau) \right] \, ,
\label{eq:poisson}
\eeq
that relates the potential to the overdensities; we have used $X$ as a shorthand notation for $DE/MG$ here.
For constant $\alpha \equiv \hat{c}_s^2$, we can build upon the discussion in \cite{Weller2003}:
 for $w>-1$ and  $\alpha \sim 1$, an initially positive DE/MG perturbation changes sign early (see Fig. 4 of \cite{Weller2003}), and then evolves with an opposite sign compared to the matter perturbation. This contribution results in a larger change in the overall density perturbation and therefore, via Eq.~\eqref{eq:poisson}, in a faster decay of the potential.  Lowering $\alpha$ causes the sign reversal to take place later. In that case, the change in the total density perturbation is smaller, so the potential decay is slower. This effect can be easily seen in the top left panel of Fig.~\ref{fig:Phi} when comparing the $(w=-0.6, \alpha = 1)$ in green with the $(w=-0.6, \alpha = 0)$ case in dark blue. 
If $w<-1$, the effect is reversed. So the potential in the $(w=-1.4, \alpha = 1)$ case decays more slowly than in the model with $(w=-1.4, \alpha = 0)$. Similar statements can be made concerning the effects of $\beta_1$ and $\beta_2$ on the perturbation evolution; for the numerical results see the upper right and lower left panels of Fig.~\ref{fig:Phi}.

The behaviour of the EDE model (bottom right panel) has to be discussed separately. For $w>-1$ the evolution of $\Phi + \Psi$ still resembles the previously discussed cases, but for $w<-1$ it is fundamentally different. The additional anisotropic stress from the DE/MG component causes the absolute value of the source term to grow instead of decaying. We can already anticipate that this will have a strong effect on the CMB spectrum and other observables. 

\subsubsection{CMB power spectrum}
\begin{figure}[tb]
\centering
\includegraphics[width=0.8\textwidth]{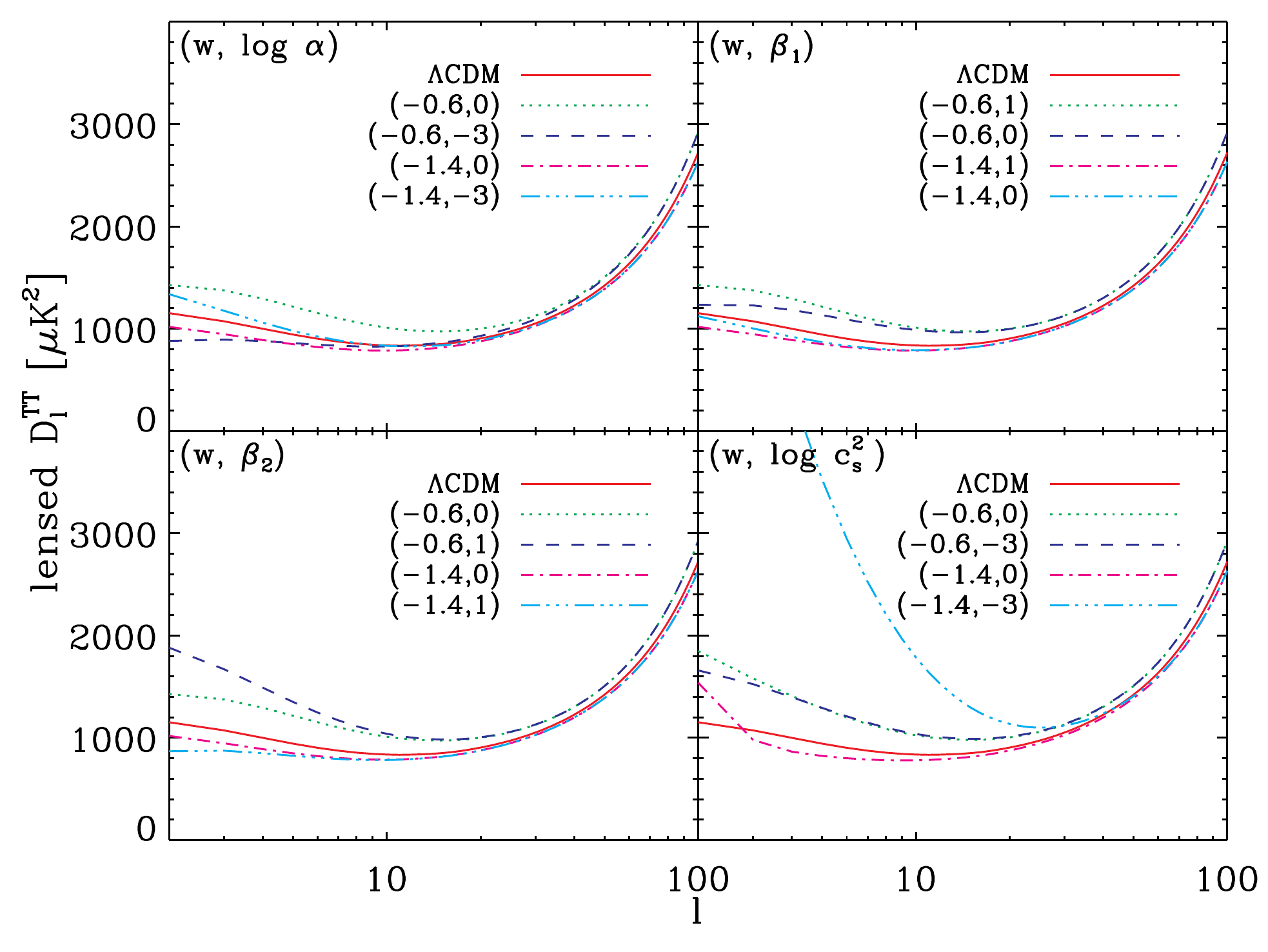}
\caption{\label{fig:ClTT} Lensed angular CMB temperature power spectrum for the different DE/MG models, where as usual $D_\ell \equiv \ell(\ell+1)C_\ell/(2 \pi)$. Parameters and colour coding are the same as in Fig. \ref{fig:Phi}.}
\end{figure}

The CMB photons propagating through the cosmic web of over- and underdensities are sensitive to the evolution of the potentials via the late ISW effect. A secondary anisotropy
\beq
\frac{\Delta T}{T} (\nhat) \equiv \Theta (\nhat) =   \int  d\tau \, e^{-\tau_{\mathrm{opt}}(z)} (\dot{\Phi}+ \dot{\Psi}) [\tau, \nhat (\tau_0 - \tau) ]
\label{eq:ISW}
\eeq
is imprinted on the CMB in the direction $\nhat$. In Fig.~\ref{fig:ClTT} we plot the lensed CMB temperature spectrum for the models we are considering. The effect of $w$ on intermediate and smaller angular scales is the well-known shift of the acoustic peaks due to the change in the angular diameter distance. Only on the largest scales, are the $C^{TT}_\ell$  sensitive to the detailed properties of the DE/MG perturbations. With the evolution of the potential already at hand, the change at low $\ell$ is straightforward to explain.
In general, a faster decay of the potentials $\Phi$ and $\Psi$ leads to a larger ISW source term (see Eq.~\ref{eq:ISW}), so that we expect an increased power in the low-$\ell$ modes. This is confirmed by the numerical results; see for example the top left panel of Figs.~\ref{fig:Phi} and \ref{fig:ClTT}: The $(w=-0.6, \alpha = 1)$ model in green shows a faster potential decay and hence a higher ISW contribution than e.g.\@\xspace the one with $(w=-0.6, \alpha = 0)$ in dark blue. 

Again we can use similar arguments to explain the change in the CMB temperature power spectrum when varying $(w,\beta_1)$ or $(w, \beta_2)$; these are shown in the second and third panels of Fig.~\ref{fig:ClTT}. In the case of the EDE model (bottom right panel), the CMB spectra for the two $w=-0.6$ cases show a higher ISW signal due to the faster potential decay (compared to $\Lambda$CDM). Thus, this model would already be in conflict with the observed CMB spectrum.
 If we however go to $w < -1$ (such as $w=-1.4$ in our plot), the low-$\ell$ $C_\ell$ are strongly enhanced. This is due to the steep growth of $|\Phi + \Psi|$ visible in the bottom right panel of Fig.~\ref{fig:Phi}. Note that both an increase (as in this case) and a faster decrease (as in the other models) of this source term lead to more power on large angular scales. This is due to the fact that the $C_\ell$ represent the variance in the temperature anisotropies at a particular angular scale, so they are sensitive to the absolute value of the potential change via the ISW effect, but not to its sign. This is not the case when considering external CMB correlations.

\subsubsection{CMB -- galaxy correlations} 
In order to isolate the ISW signal from the primary CMB anisotropies, it is possible to cross-correlate the CMB with a tracer of the LSS, such as galaxy catalogues \cite{Crittenden1995}.
Galaxies trace the matter overdensities $\delta_m(\nhat,z)$, which are in turn related to $\Phi$ via the linear Poisson equation \eqref{eq:poisson}, up to some bias  $b$. As we are working in linear perturbation theory, we can generally assume the bias to be linear and scale-independent, so that we can write
\beq
\delta_g(\nhat,z) = b(\nhat,z) \delta_m(\nhat,z) \, .
\eeq
As a simplified model for the linear bias, we adopt the redshift dependence
\beq
b_i(z) = 1+\frac{b_0^i -1 }{[ D(z) ]     ^{\gamma_i}}
\label{bias}
\eeq
as in \cite{Giannantonio2013}, where $D(z)$ is the linear growth factor and $\gamma_i =2$ is a reasonable choice in the case of flux-limited surveys \cite{1998MNRAS.299...95M}. We will consider the $\{b_0^i\}$  as free nuisance parameters for every galaxy catalogue we use.  
When cross-correlating the galaxy overdensity $\delta_g(\nhat,z)$ with the temperature anisotropies in the CMB, it is necessary to project the former onto the sphere via
\beq
\delta_{g_i} (\nhat) = \int dz \,  \delta_{g_i}(\nhat,z) \htwo \phi_i(z) \, ,
\eeq
where $\phi_i(z)$ is the normalised visibility function of the respective galaxy survey. The
angular temperature-galaxy cross-correlation functions (CCFs) are then defined as

\beq
w^{Tg_i} (\vartheta) = \langle \Theta (\nhat_1) \delta_{g_i}(\nhat_2) \rangle \, ,
\eeq
where the average is over all pairs with angular separation $\vartheta = \arccos(\nhat_1\cdot \nhat_2)$.
We compute the theoretical predictions for the CCFs with a modified version of \textsc{Camb} that also contains the DE/MG models as described above. For this purpose, we first calculate the auto- and cross-spectra in harmonic space and from these the corresponding correlation functions via the standard Legendre transformation, see \cite{Giannantonio2012, Giannantonio2013} for details.

  \begin{figure}[tb]
  \centering
  \includegraphics[width=0.8\textwidth]{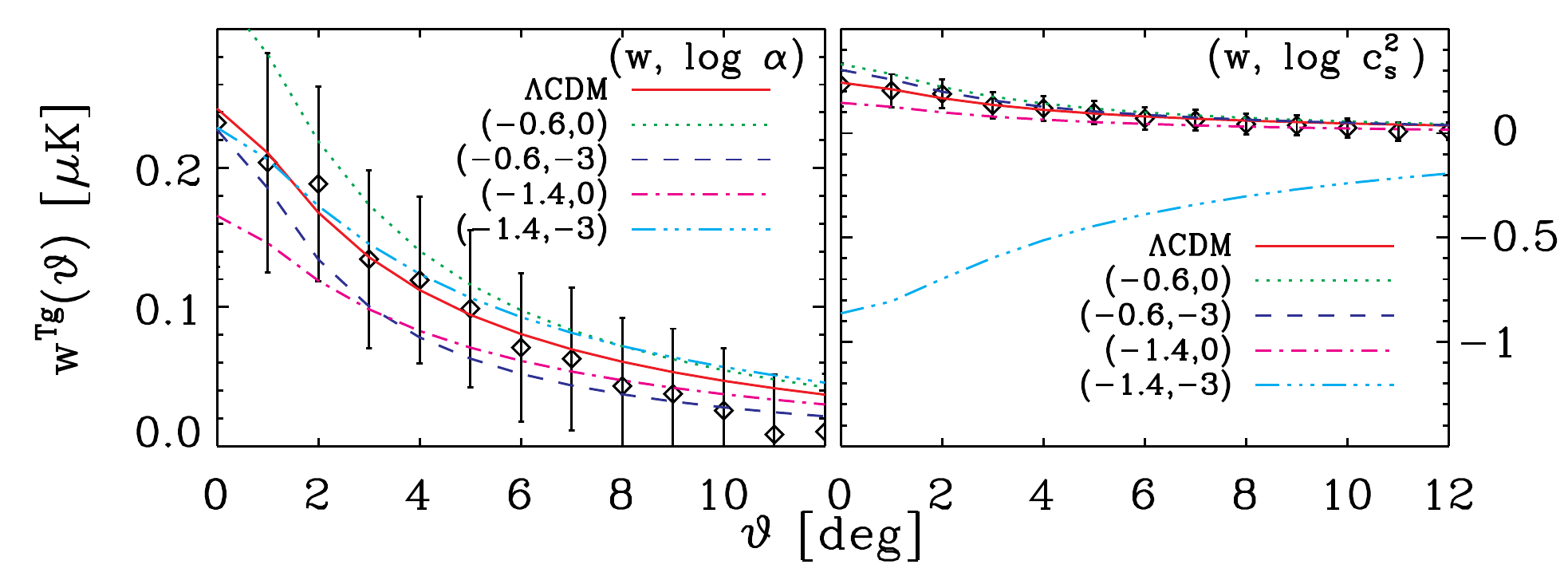}
  \caption{\label{fig:Wth_T_NVSS} Theoretical predictions and observational data for the $Tg$-correlation functions (CFs) (\Planck -- NVSS correlation). Colour coding is the same as in the previous figures, but we have omitted the panels with $\beta_1$ and $\beta_2$ because these parameters only have a small effect on the $Tg$-CFs. Note the change in vertical scale between the left and right panel.}
  \end{figure}

 In Fig.~\ref{fig:Wth_T_NVSS} we show the effect of our DE/MG parameterisation on these correlations. In addition to the theoretical predictions, we also plot the measured temperature-galaxy correlation of the \Planck CMB temperature maps with the NRAO VLA Sky Survey (NVSS) radio galaxy catalogue \cite{Condon1998}, which is one example from our full data set, as described in Section~\ref{sec:dataset} below.
  Generally speaking, we expect a higher $Tg$-correlation when there is a faster potential decay and hence a stronger ISW signal. Again, this expectation is confirmed by our numerical results. As before, the $w<-1$ case of the EDE model has to be interpreted separately. In that  scenario, the (in absolute value) growing potentials lead to a suppression of the $Tg$-correlation. For extreme cases with very low $w$ and $c_s^2$, this even causes a $w^{Tg} < 0$, i.e. an anticorrelation; see the right panel of Fig.~\ref{fig:Wth_T_NVSS}. We can easily anticipate that this parameter region will be ruled out by comparison with the data.
 
 We  also compute  galaxy-galaxy auto-correlation functions (ACFs) and CCFs. These are defined as
 \beq
 w^{g_i g_j} (\vartheta) = \langle \delta_{g_i} (\nhat_1) \delta_{g_j}(\nhat_2) \rangle
 \eeq
 and are calculated in a similar manner as the temperature-galaxy CCFs described above. 
The $gg$-correlations are less sensitive to the details of DE/MG perturbations and so we do not show them explicitly here, but they are important to constrain the values of the bias parameters.

\begin{figure}[tbp]
 \centering 
\includegraphics[width=0.8\textwidth]{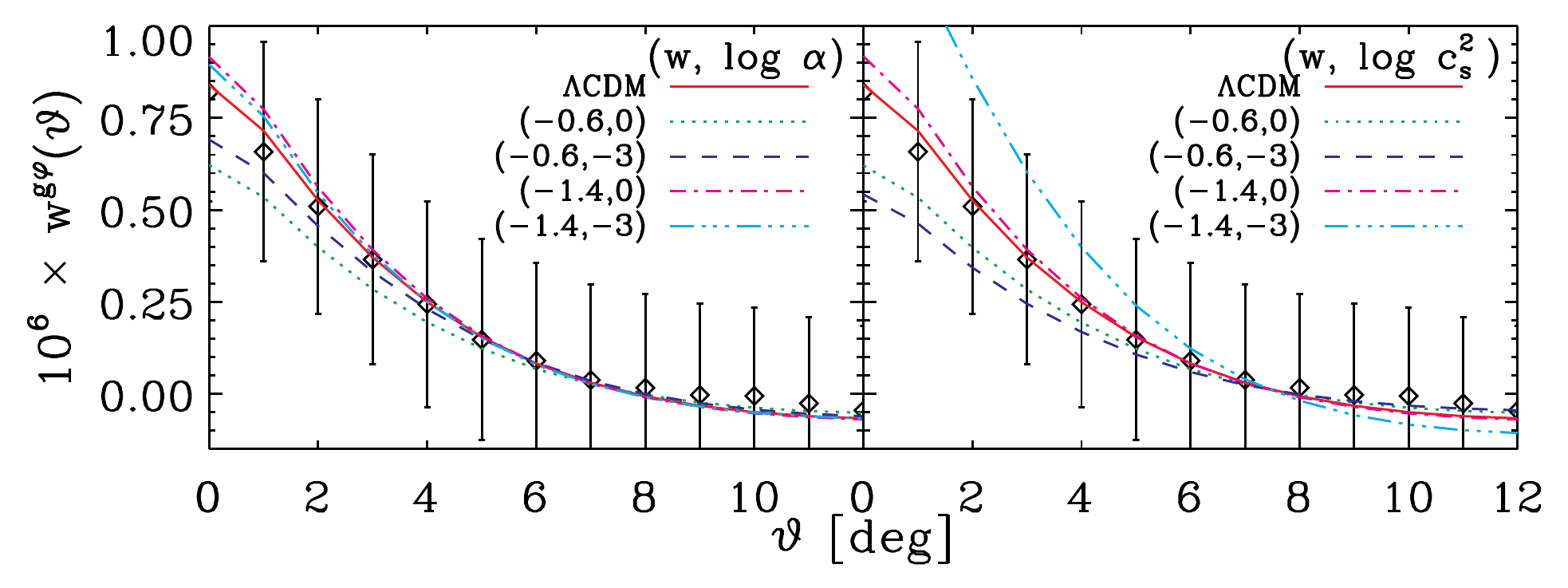}
 \caption{\label{fig:Wth_Phi_LRG}
 Theoretical predictions and observational data for the $\varphi g$-CFs (\Planck lensing potential and SDSS LRGs). Panels and colour coding are the same as in Fig.~\ref{fig:Wth_T_NVSS}.}
 \end{figure}

  Finally, we also take into account the correlations between the CMB lensing potential reconstructed from the \Planck temperature maps \cite{PlanckXVII} and the galaxy catalogues. This additional probe has recently been introduced into the data set by \cite{Giannantonio2013b}. The lensing potential is computed from \cite{Lewis2006}
 \beq
 \varphi(\nhat) = - \int_{0}^{\chi_*} d\chi \, \left[\frac{\chi_* - \chi}{\chi_* \chi}  \right] \, \left[\Phi + \Psi \right] \left(\chi \nhat, \tau_0 - \chi \right) \, ,
 \eeq
with $\chi$ being the conformal distance and the asterisk denoting the time of the last scattering. We can thus define the corresponding CMB lensing--galaxy correlation functions (CFs) as
\beq
w^{\varphi g_j} (\vartheta) = \langle \varphi (\nhat_1) \delta_{g_j}(\nhat_2) \rangle \, .
\eeq
In Fig.~\ref{fig:Wth_Phi_LRG} we show the correlation between the lensing potential $\varphi$ and the galaxies from the Sloan Digital Sky Survey Data Release 8 Luminous Red Galaxies (SDSS-DR8-LRG) sample from \cite{Ross2011}; this is again one example from our full data set. These lensing correlations have smaller statistical uncertainty than the $Tg$ ones, but their sensitivity to the  properties of DE/MG perturbation evolution is lower, since the CMB lensing kernel spans a wide redshift range and is only partially affected by the DE/MG phenomenology at late times.

\section{Data set}
\label{sec:dataset}
We constrain the background cosmology with the CMB temperature power spectrum as measured by the \Planck satellite \cite{Planck2013,Planck2013XV}, including the low-$\ell$ CMB polarization from the \WMAP 9-year results (WP) \cite{Bennett2013a}. As a low-redshift probe of the expansion history we use the BAO measurements at three different redshifts from the 6dF Galaxy survey \cite{Beutler2011}, the Sloan Digital Sky Survey Data Release 7 (SDSS DR7) analysed by \cite{Padmanabhan2012} and the SDSS DR9 Baryonic Oscillation Spectroscopic Survey (BOSS) \cite{Anderson2013}. We have also tested the stability of our results to the inclusion of the Union 2.1 \cite{Suzuki2012} and Supernovae Legacy Survey (SNLS) \cite{Conley2011} SNe Ia samples, but do not use them in our final analysis. 

Our additional ISW/LSS/CMB lensing data set is built from the \Planck CMB temperature and lensing maps, and six galaxy catalogues: the infrared galaxies from 2MASS \cite{Jarrett2000} (median redshift $z_m \sim 0.1$), the SDSS DR8 main galaxy sample \cite{Aihara2011} with \mbox{$z_m \sim 0.3$} and the SDSS photometric LRGs from the DR8-CMASS sample \cite{Ross2011} ($z_m \sim 0.5$). It also contains the NVSS radio galaxies \cite{Condon1998},  the X-ray background from the High Energy Astronomy Observatory (HEAO) \cite{Boldt1980} and finally the SDSS DR6 quasars (QSO) by \cite{Richards2008}. Catalogues, where we do not explicitly state $z_m$, have a broader distribution in redshift. This data set was first compiled in \cite{Giannantonio2008} and later updated in \cite{Giannantonio2012, Giannantonio2013} and \cite{Giannantonio2013b}, where the \Planck lensing correlations were included.
All maps are pixelated with \textsc{HEALPix} \cite{Gorski2005} at a resolution of $N_{\mathrm{side}} = 64$; the corresponding pixel size of \mbox{$50\,$arcmin} is sufficient as we are only interested in the large angular scales. The two-point CFs between the maps are then computed with the estimator described in \cite{Giannantonio2013}. The full data set consists of six $Tg$-, six $\varphi g$- and 21 $gg$-CFs. From these we discard the autocorrelations of the NVSS and QSO catalogues due to the high level of residual systematic contamination (see e.g.  \cite{Giannantonio2012,Giannantonio2013,2013PASP..125..705P, 2013MNRAS.435.1857L}). The CCFs with these catalogues are kept as they are less affected by potential systematics. The remaining 31 CFs constitute our ISW/LSS/CMB lensing data set, which corresponds to the `fair' sample of \cite{Giannantonio2013}, but contains the update from WMAP7 to \Planck and the lensing correlations (as in \cite{Giannantonio2013b}). The full covariance matrix of the data set is estimated from $10,000$ Monte Carlo realizations assuming a fiducial \LCDM model. Several possible sources of noise and systematics effects were tested; see \cite{Giannantonio2012, Giannantonio2013} for details. 

The data that are most sensitive to the DE/MG perturbation evolution are the $Tg$-CFs, sensitive to $\dot{\Phi} + \dot{\Psi}$ via the ISW effect. Unfortunately, they are also the data with the largest statistical uncertainty. The $gg$-CFs have higher precision, but their sensitivity to $\Phi$ via the Poisson equation is lower. The $\varphi g$-CFs also probe the evolution of the potential via the lensing source term $\Phi+\Psi$; their sensitivity to DE/MG and their uncertainty are intermediate.

\section{Results and discussion}
\label{sec:mcmcresults}
We constrain the DE/MG perturbation evolution jointly with the baseline cosmological parameters, using an extended version of \textsc{CosmoMC} \cite{Lewis2002}. The version of this Markov Chain Monte Carlo (MCMC) code that was released together with the 2013 \Planck data uses 14 nuisance parameters for the foreground cleaning in addition to the conventional six \LCDM cosmological parameters. For the ISW/LSS/CMB lensing data set, we add nine further nuisance parameters: one bias parameter for each galaxy catalogue (see Eq.~\ref{bias}) and three stellar contamination fractions (as defined in \cite{Giannantonio2013}) for the SDSS-based galaxy samples. Our full set of both theory and nuisance parameters is summarised in Table~\ref{tab:parameters}.

\begin{table}[tbp]
\centering
\begin{tabular}{|c||c|c|c|}
\hline
  & Parameter & Description & Prior \\

\hline \hline
\textbf{Baseline} & $\Omega_b h^2$& physical baryon density & $[0.005,0.1]$\\
											& $\Omega_c h^2$ & physical CDM density & $[0.001,0.99]$\\
											& $100  \times \theta^* $ & sound horizon at last scattering & $[0.5,10]$ \\
											& $\tau_{opt}$ & reionisation optical depth & $[0.01,0.8]$\\
											& $\ln(10^{10} A_s)$ & primordial perturbation amplitude  & $[2.7,4]$ \\
											& $n_s$ & scalar spectral index  				& $[0.9,1.1]$ \\
  \hline
  \textbf{DE/MG}	& $w$ & background equation of state & $[-3,-0.3]$\\ 
  \hdashline
  $w\Gamma$ model					& $\log_{10}(\alpha)$ & DE/MG perturbation parameters & $[-3,0]$  \\
  						& $\beta_1$ 			   &	(entropy perturbation model)		& $[0,2]$\\
  						& $\beta_2$					&		& $[0,15]$ \\
  						 \hdashline
  $w\Pi^s$ model & $\log_{10}(c_s^2)$  		&	elastic DE sound speed & $[-3,0]$ \\						 			
  \hline
  \textbf{Nuisance} & $b_0^i$ & 6 bias parameters (1 per catalogue)& $[0,3]$\\
  \textbf{(correlation data)}			& $\kappa_i$& 3 SDSS stellar contamination fractions& $[0,0.1]$ \\
  \hline
  \textbf{\Planck nuisance} & 14 params. & foreground nuisance parameters  & (see \cite{Planck2013XV}) \\ 
\hline
\end{tabular}
\caption{\label{tab:parameters} Summary of the parameters and their prior ranges for the MCMC runs. }
\end{table}

\subsection{Background equation of state}
\begin{figure}[tbp]
\begin{center}
\subfigure{\includegraphics[width=0.40\textwidth]{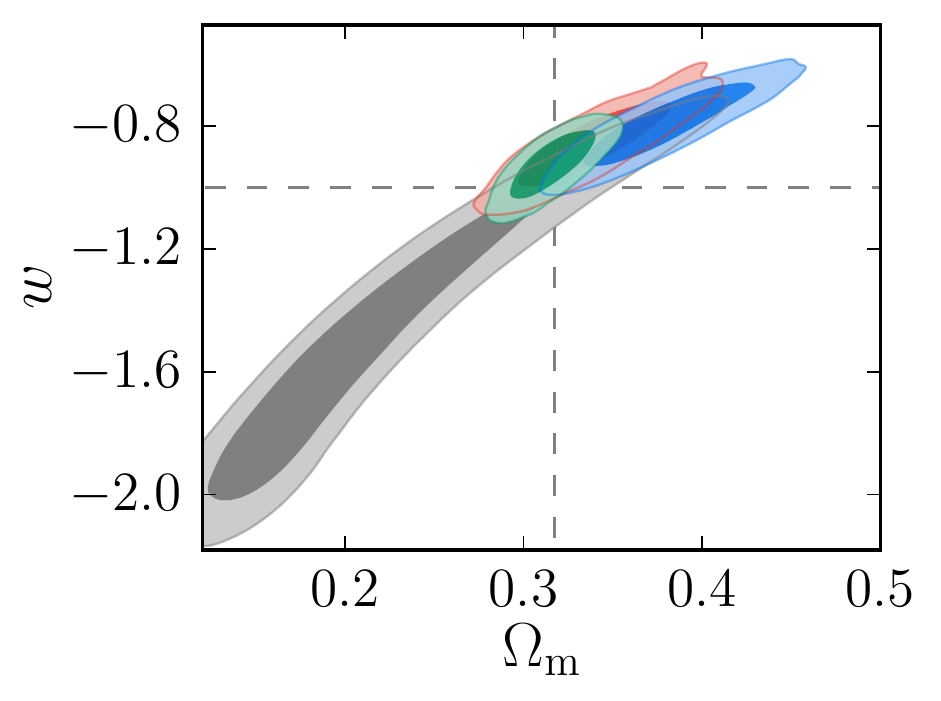}}\hfill
\subfigure{\includegraphics[width=0.59\textwidth]{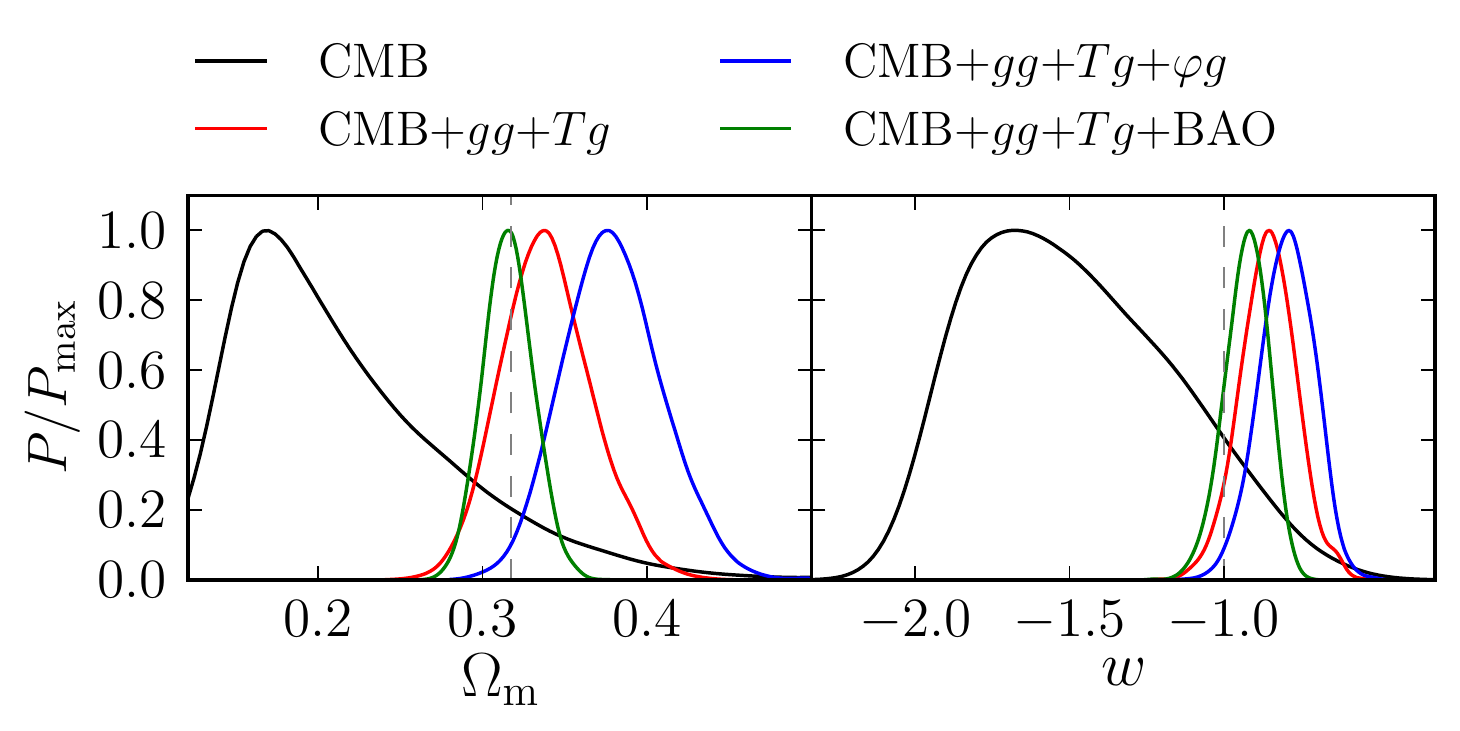}}
\caption{\label{fig:w_omm_1D_2D} Results for the $w$CDM model.  \textbf{Left panel}: Constraint on the background equation of state $w$ versus the matter content $\Omega_m$ when combining the CMB with different parts of the cross-correlation and BAO data sets. \textbf{Right panel}: Marginalised posterior distributions of  $\Omega_m$ and $w$ with the same data sets. We have highlighted $\Omega_m = 0.3175$ (\Planck $\Lambda$CDM best fit) and $w=-1$.}
\end{center}
\end{figure}

We first present constraints on a $w$CDM cosmology (corresponding to a minimally coupled quintessence with $\alpha =1$, $\beta_1 = 1$ and $\beta_2=0$) while leaving the perturbation evolution unmodified. When using only the CMB, the well-known geometric degeneracy makes it impossible to make meaningful statements about $w$. In the original \Planck analysis \cite{Planck2013}, this degeneracy is broken by adding (amongst others) the BAO data, giving $w = -1.13^{+0.24}_{-0.25}$ at $95\%$ confidence level (CL). Using the $Tg$- and $gg$-CFs instead of the BAOs we find in agreement with \cite{Giannantonio2013}:
\beq
w = -0.86^{+0.19}_{-0.18} \qquad (95\% \htwo \text{CL; CMB}+gg+Tg) \, .
\eeq
When adding the $\varphi g$-CFs, i.e.\@\xspace using the full ISW/LSS/CMB-lensing data set, we find
\beq
w = -0.80 ^{+0.16}_{-0.18} \qquad (95\% \htwo \text{CL; CMB}+gg+Tg+\varphi g) \, ,
\eeq
so the $\varphi g$-CFs shift the constraint for $w$ to higher values and hence results in a slight ($\sim 2 \sigma$) preference for $w>-1$. 
We decided not to use the $\varphi g$-CFs when including the BAO data, as the former prefer a higher $w$ than the latter; so combining them would potentially have led to unnaturally tight constraints due to the tension between the data sets. With the inclusion of BAO data we find
\beq
w = -0.93 \pm 0.14  \qquad (95\% \htwo \text{CL; CMB}+gg+Tg+\text{BAO}) \, ,
\eeq
which is a considerable improvement of the constraint on $w$ and fully consistent with $\Lambda$CDM. 
In Fig.~\ref{fig:w_omm_1D_2D} we show the marginalised 1D posteriors of $w$ and $\Omega_m$ as well as the 2D likelihood contours for the different combinations of data sets.

\subsection{Entropy perturbation model: $w$ and $\alpha$}
We then open up the space of DE/MG perturbation evolution parameters, starting with the entropy perturbation model. As a first step, we set $\beta_1 =1$ and $\beta_2 = 0$ and only use $\alpha$ in addition to the previously discussed parameters. In the fluid description, this parameter corresponds to the sound speed $\hat{c}_s^2$ of DE perturbations; many quintessence models can already be described by $w$ and $\alpha$ at the level of linear perturbations. On physical grounds, we should therefore impose $0 \leq \alpha \leq 1$. If one however directly uses $\alpha$ as a parameter with this prior range, one indirectly penalises small values of $\alpha$ as only a very small prior volume is occupied by them. For that reason, $\log_{10}(\alpha)$ is a more natural parameter choice. By using a flat prior in log-space, we do not impose any a priori assumption concerning the order of magnitude \nolinebreak[4] of \nolinebreak[4] $\alpha$. 

\begin{figure}[tbp]
\begin{center}
     \includegraphics[width=0.8\textwidth]{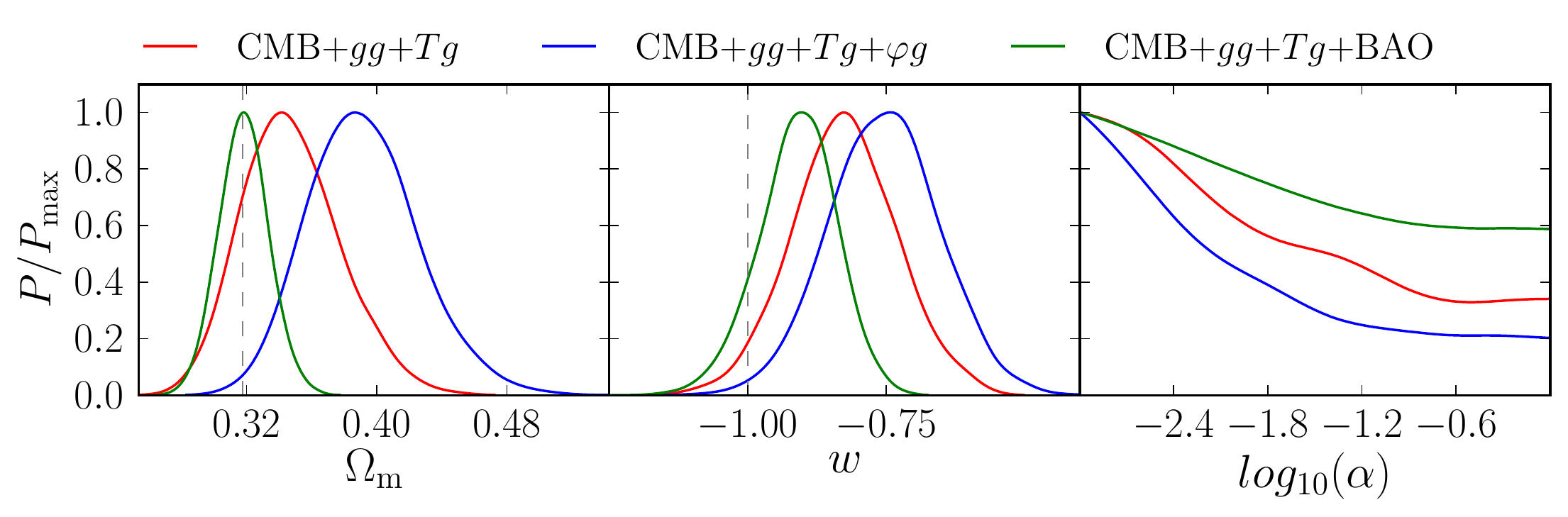}\\
     \includegraphics[width=0.7\textwidth]{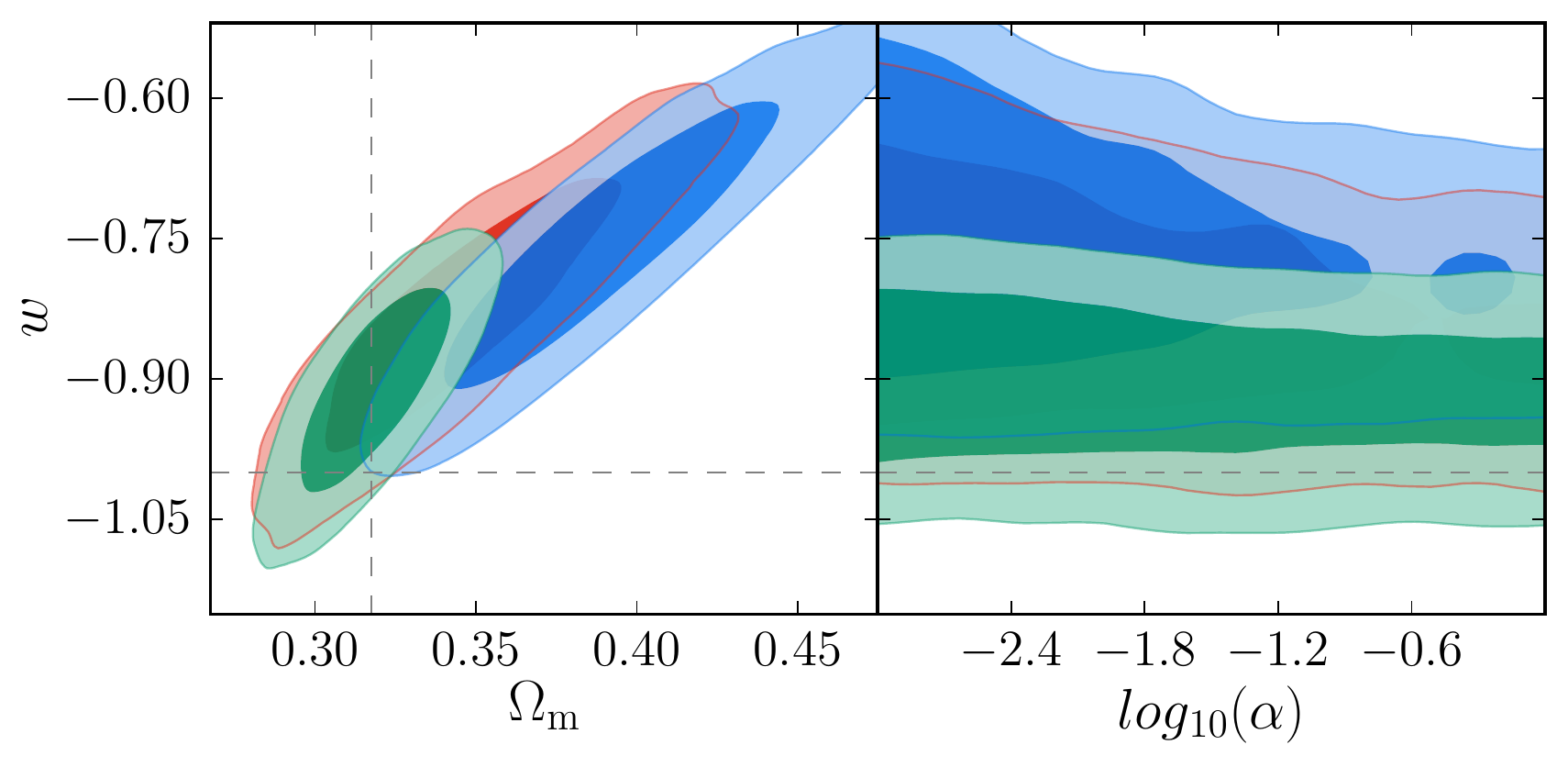}
 \caption{\label{fig:w_alpha_1D_2D} Results for the simplified entropy perturbation model.  \textbf{Upper panels:} Marginalised 1D posteriors of $\Omega_m$, $w$ (background) and $\log_{10}(\alpha) $  when using different combinations of data. \textbf{Lower panels:} Marginalised 2D likelihood contours in the $w$-$\Omega_m$ and $w$-$\log_{10}(\alpha)$ plane.  Except for the `CMB only' case, which we do not show here, the data sets used and the colour coding are the same as in Fig.~\ref{fig:w_omm_1D_2D} in all panels.}
 \end{center}
 \end{figure}

We have found that the effect of $\alpha$ on our observables `saturates' for $\log_{10}(\alpha) \ll -1$, so that a reasonable choice for the prior in log space is to impose $\log_{10}(\alpha) > -3$; with this we probe sufficiently small values of $\log_{10}(\alpha)$.
With this prior range, we find
\beq
w = -0.83^{+0.19}_{-0.18} \qquad (95\% \htwo \text{CL; CMB}+gg+Tg ) \, 
\eeq
from the CMB and the $Tg$-CFs. Again, this constraint shifts to higher values of $w$ by including the $\varphi g$-CFs, in which case we have
\beq
w = -0.75^{+0.19}_{-0.20} \qquad (95\% \htwo \text{CL; CMB}+gg+Tg+\varphi g ) \, ,
\eeq
where a $\sim 2 \sigma$ tension with $\Lambda$ appears.
If we use the BAO data set instead, the result is
\beq
w =-0.91^{+0.13}_{-0.15} \qquad (95\% \htwo \text{CL; CMB}+gg+Tg+\text{BAO} ) \, .
\eeq
We show in Fig.~\ref{fig:w_alpha_1D_2D} the resulting marginalised one-dimensional posterior distributions of $\Omega_m$, $w$ and $\log_{10}(\alpha)$ and the corresponding two-dimensional likelihood contours in the $w$--$\Omega_m$ and $w$--$\log_{10}(\alpha)$ plane when using these different combinations of data.
In the cases without the BAOs we see a tendency towards $\log_{10}(\alpha) < 0$, i.e.\@\xspace a clustering in the DE component:
\begin{align}
\log_{10}(\alpha) &< - 1.45 \qquad (68\% \htwo \text{CL; CMB}+gg+Tg ) \, ; \\
\log_{10}(\alpha) &< -1.69 \qquad (68\% \htwo \text{CL; CMB}+gg+Tg+\varphi g ) \, ;
\end{align}
the significance is however well below the $2 \sigma$ level. If the BAOs are included, these indications disappear because $w$ is constrained to a region rather close around $w = -1$. In that case, models with different $\log_{10}(\alpha)$ are completely indistinguishable as there is almost no effect on the observables. This is however not a problem of our particular parameterisation or data sets but a general challenge to DE/MG science: if the equation of state is very close to $w=-1$, both DE and MG simply mimic a cosmological constant and it is not possible to distinguish between different models.

There is another interesting feature of the results of this section: the constraints on $w$ have all moved to slightly higher values of $w$ in comparison to the previous ones obtained in the `background only' $w$CDM model, which is due to the degeneracy in the effects of $w$ and $\alpha$ on the CMB. A higher $w$ causes a stronger ISW effect (see Fig.~\ref{fig:ClTT}) and hence also higher $Tg$-correlation (see Fig.~\ref{fig:Wth_T_NVSS}). In the $w$CDM model, a $w$ significantly larger than $-1$ will therefore be in tension with the data. If we however use $\log_{10}(\alpha)$ as an additional parameter, the effect can be compensated for by lowering $\log_{10}(\alpha)$, so a model with a slightly higher $w$ can still fit the data well. 

This statement can be generalised to the other models of DE/MG perturbation evolution that we will examine in the following sections: we will get different results for $w$ depending on how we treat the perturbative behaviour of the dark sector. Therefore the constraint on $w$ is model-dependent.
We can push this argument even further: the constraint on $w$ not only changes between the different models, but also depends on the prior that we impose for DE/MG parameters like $\log_{10}(\alpha)$. To show this, we have performed the parameter estimation for the `CMB+$gg$+$Tg$' data combination with different priors on $\log_{10}(\alpha)$; the results are summarised in Table~\ref{tab:w_alpha_prior}.
\begin{table}[tbp]
\small
\centering
\begin{tabular}{|c||c|c|c|c|}
\hline 
\textbf{Prior on $\log_{10}(\alpha)$} &  $[-4,0]$ & $[-3,0]$ & $[-2,0]$ & $[-1,0]$\\
\hline
 \textbf{Constraint on $w$} (95\% CL)& $-0.81 \pm 0.19$ &$-0.83^{+0.19}_{-0.18}$ & $ -0.85 \pm 0.18$ & $-0.86^{+0.17}_{-0.18}$\\
\hline
\end{tabular}
\caption{Constraint on $w$ depending on the prior of $\log_{10}(\alpha)$ for the `CMB+$gg$+$Tg$' data combination.}
\label{tab:w_alpha_prior}
\end{table}

\subsection{Full entropy perturbation model}
\begin{figure}[tbp]
\centering
\begin{center}
\includegraphics[width=\textwidth]{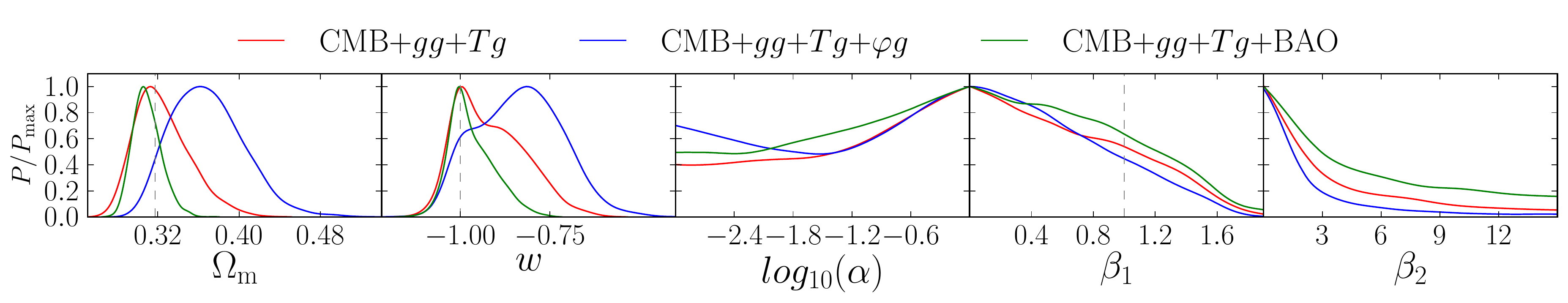}
\includegraphics[width=0.7\textwidth]{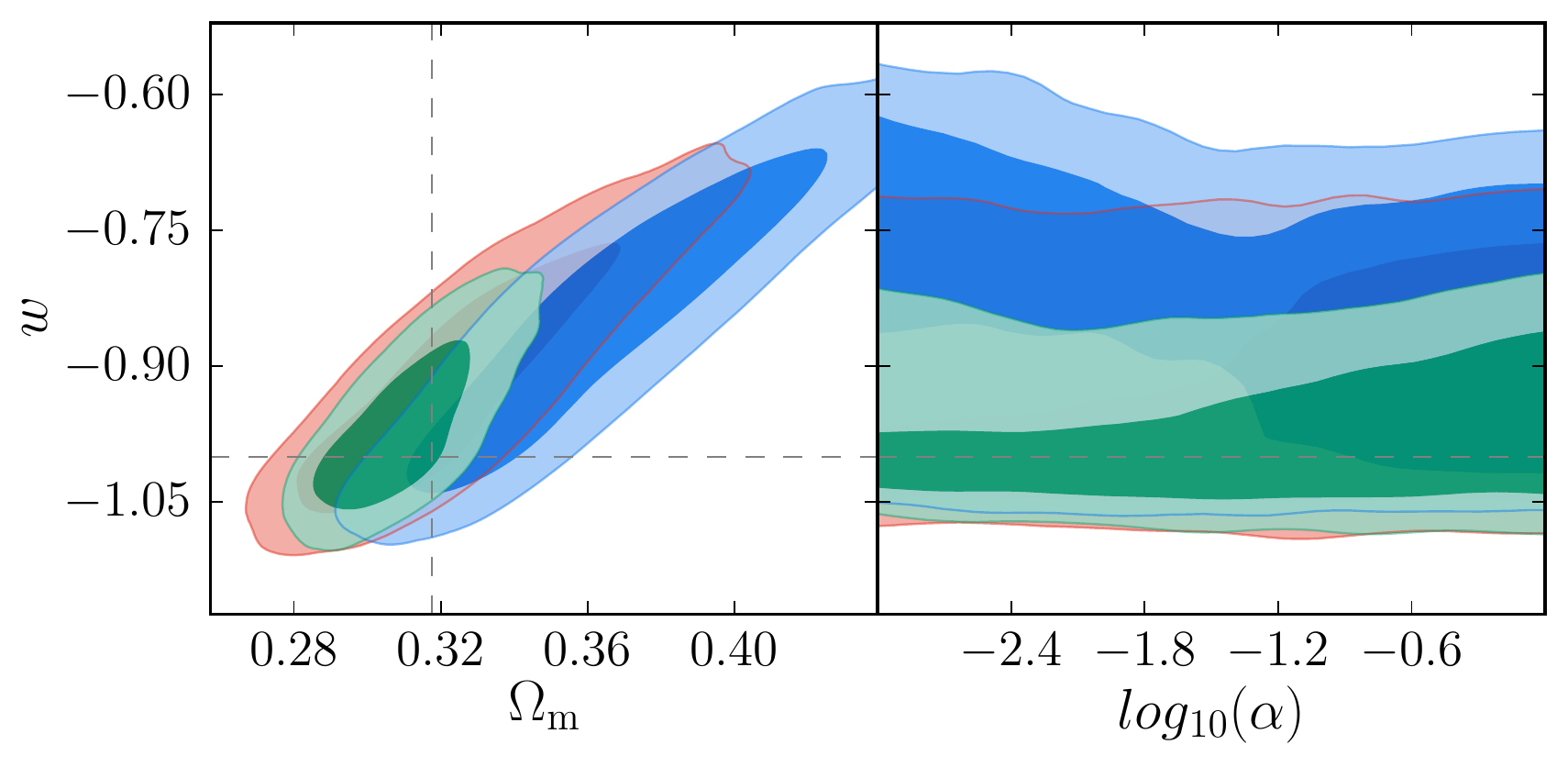}
\includegraphics[width=0.7\textwidth]{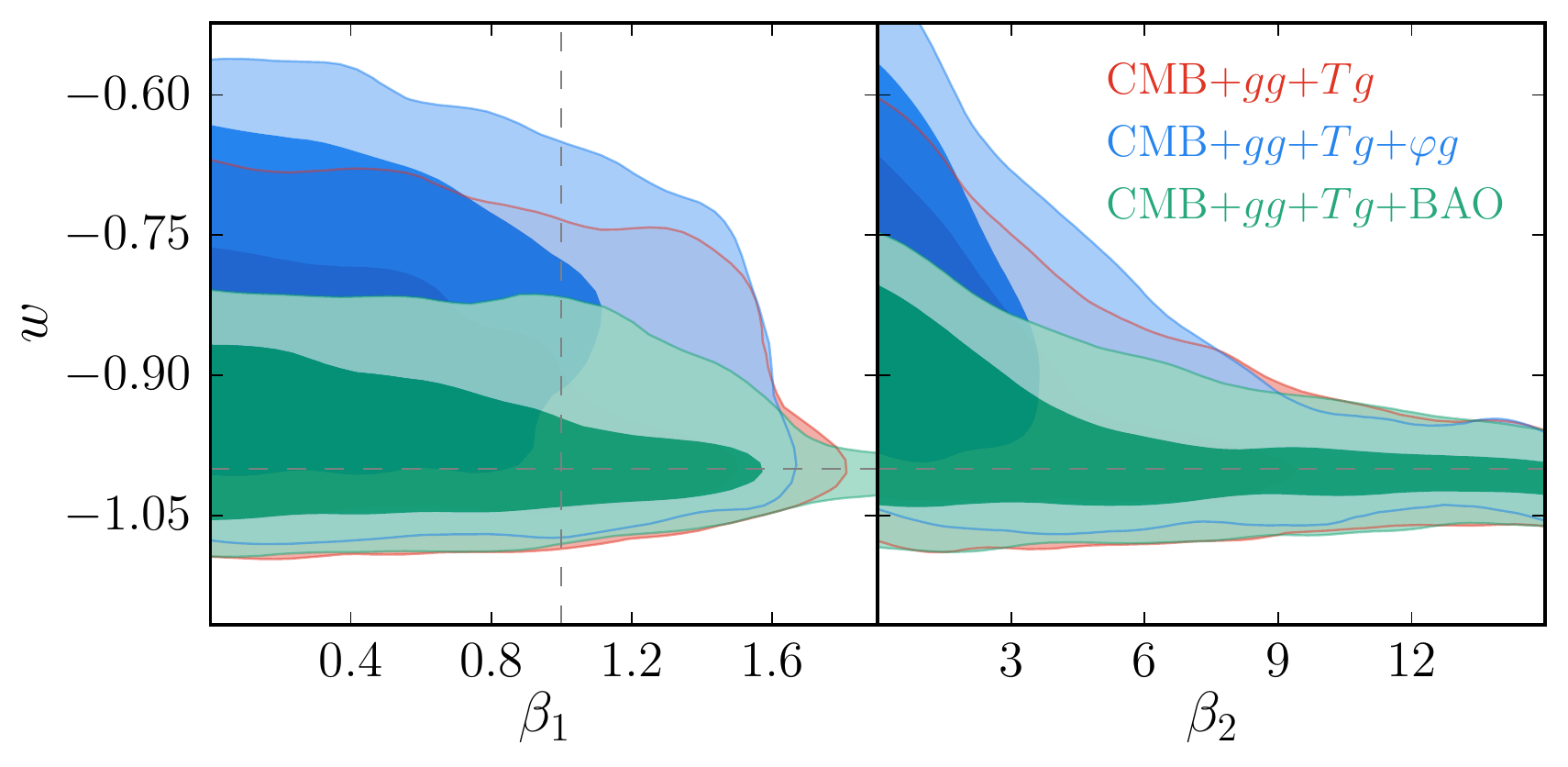}
\end{center}
\caption{\label{fig:wab12_1D_2D}  Results on the full entropy perturbation model.  \textbf{Upper panels:} Marginalised one-dimensional posteriors of the relevant parameters in the full entropy perturbation model: $\Omega_m$ and $w$ (background parameters), $\log_{10}(\alpha)$, $\beta_1$ and $\beta_2$ (perturbation evolution parameters). 
\textbf{Middle and lower panels:} Constraints on $w$ versus $\Omega_m$, $\log_{10}(\alpha)$, $\beta_1$ and $\beta_2$ for the full entropy perturbation model. The data sets used and the colour coding are the same as in Fig.~\ref{fig:w_alpha_1D_2D}. \vspace{0.5cm}}
\end{figure}

We now use the full parameter set $\{w, \alpha, \beta_1, \beta_2\}$ of the entropy perturbation model. For $\beta_1$ and $\beta_2$ we use relatively wide flat priors, which are chosen so that very high values of $\beta_1$ and $\beta_2$ will cause the theoretical prediction to disagree significantly with the data unless $w \simeq -1$.  The results obtained with the different combinations of data are shown in Table~\ref{tab:entropy_result} and Fig.~\ref{fig:wab12_1D_2D}.

We note that again the result for $w$ has moved significantly: now all data combinations give a value closer to $w=-1$. This supports our earlier finding that the constraints on this parameter are indeed model-dependent. Regarding the parameter $\log_{10}(\alpha)$, we note that the mild tendency towards low values that we have found in the $w$-$\log_{10}(\alpha)$ model has disappeared; now the posterior of the latter is relatively flat with a tendency towards higher values. The situation is, however, different for the parameters $\beta_1$ and $\beta_2$. Here we obtain significant upper limits, although these are only meaningful if $w \neq -1$. Including the BAO data sets again enforces $w \approx -1$, so in that case there are no constraints on these parameters.

 The bottom line for this very general model is that we are indeed able to rule out some regions of the full parameter space. Nevertheless, current data still leaves considerable freedom for the perturbation evolution in this model, especially if $w \approx -1$.  We reiterate that in the latter case the perturbations only play a minor role and vanish completely if $w = -1$; for this reason it is not possible to make any meaningful statement about the other perturbation evolution parameters in this limit.

\begin{table}[tbp]
\small
\centering
\begin{tabular}{|c||ccc|}
\hline
 & CMB+$gg$+$Tg$ & CMB+$gg$+$Tg$+$\varphi g$ & CMB+$gg$+$Tg$+BAO \\
 
\hline \hline

$w$ & $-0.92^{+0.20}_{-0.16}$ (95\%) & $-0.84 \pm 0.22$ (95\%) & $-0.97^{+0.14}_{-0.11}$ (95\%) \\[1ex]

$\log_{10}(\alpha)$ & > -1.69 (68\%) & -- & > -1.74 (68\%)\\[1ex]

$\beta_1$			&	< 1.45	(95\%) &   < 1.37  (95\%)    &					< 0.90 (68\%)									\\[1ex]

$\beta_2$ & < 12.05 (95\%)& < 9.79 (95\%) & < 6.56 (68\%)\\
\hline
\end{tabular}
\caption{Mean value and confidence regions of the parameters of the full entropy perturbation model. We quote 95\% CL limits wherever possible; otherwise we give 68\% CL intervals.}
\label{tab:entropy_result}
\end{table}

\subsection{Anisotropic stress model}
\begin{figure}[tbp]
\begin{center}
\includegraphics[width=0.8\textwidth]{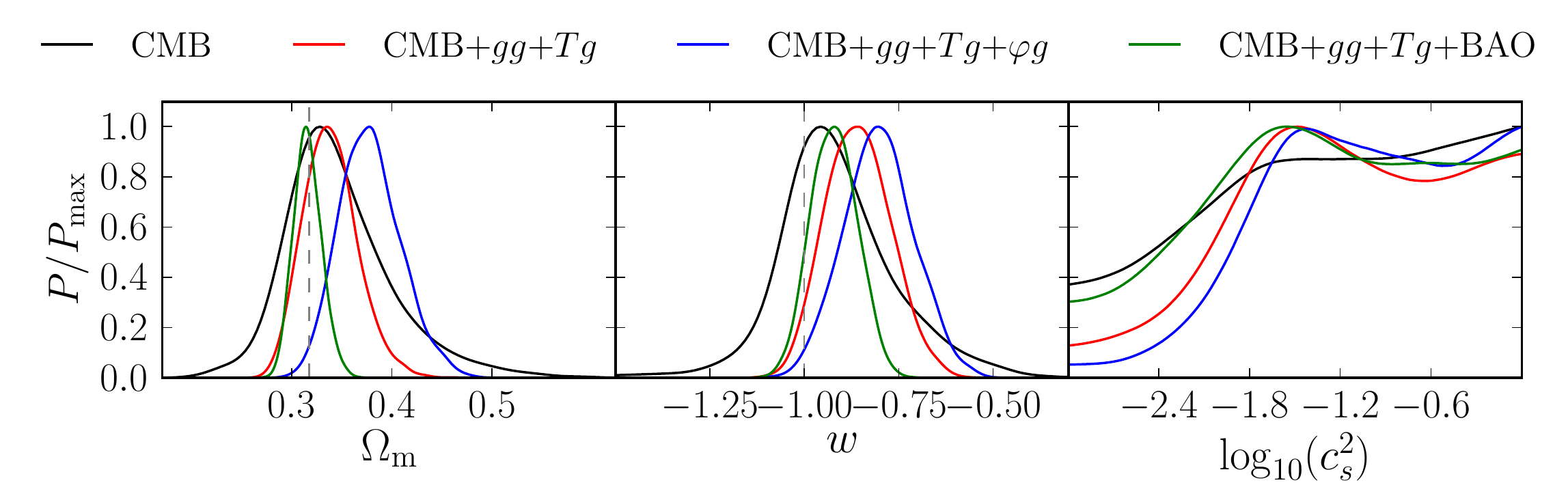}\\
\includegraphics[width=0.7\textwidth]{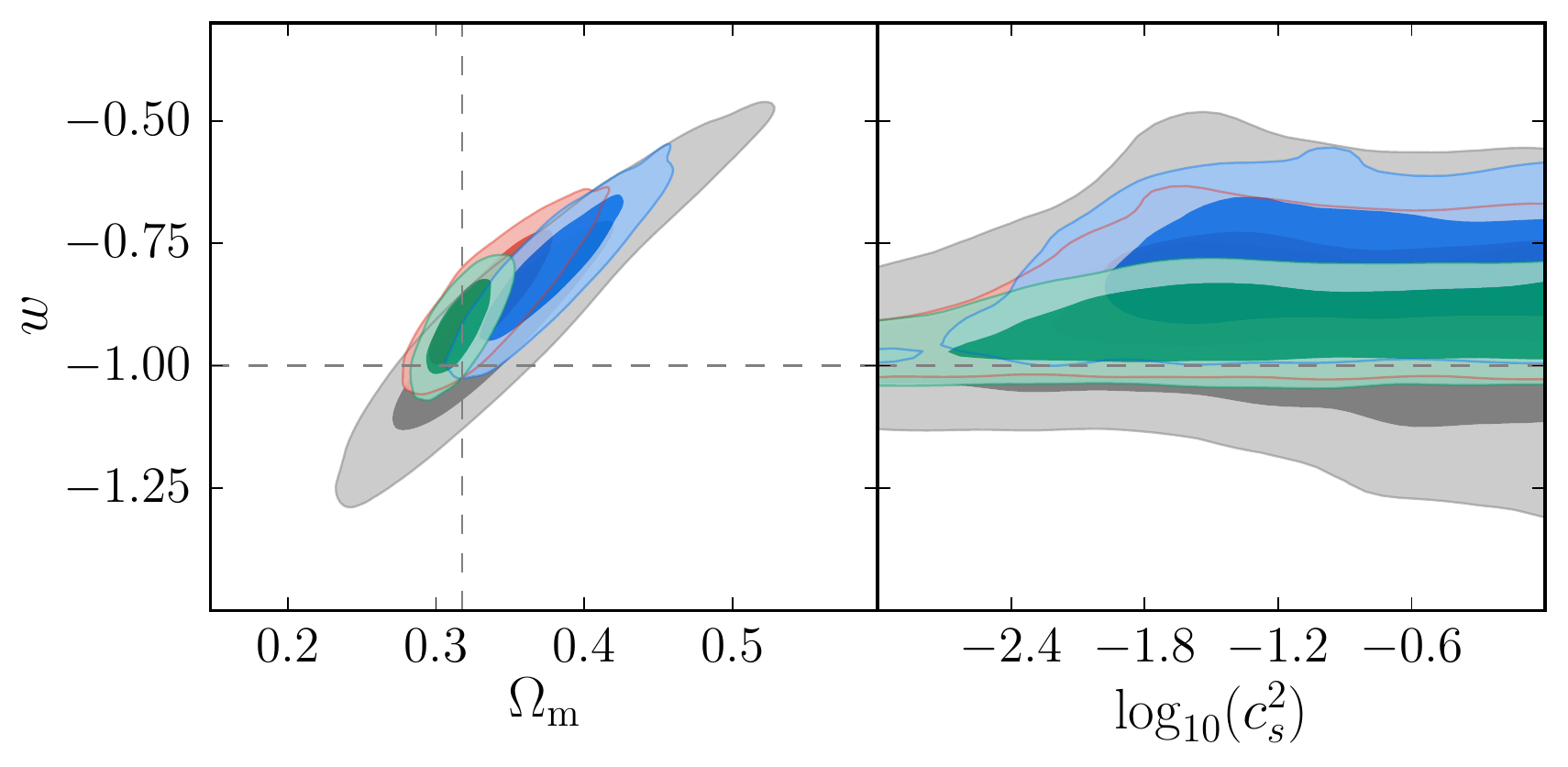}
\caption{\label{fig:w_cs2_1D_2D} Results on the anisotropic stress model. \textbf{Upper panels:} Marginalised 1D posteriors of $\Omega_m$, $w$ (background parameters) and $\log_{10}(c_s^2)$  from the CMB only and in different combinations with the cross-correlation and BAO data sets. \textbf{Lower panels:} Marginalised 2D likelihood contours in the $w$-$\Omega_m$ and $w$-$\log_{10}(c_s^2)$ plane. Except from the `CMB only' case (black/grey), the data sets used and the colour coding are the same as in the previous figures. }
\end{center}
\end{figure}

We have also tested the anisotropic stress model assuming a flat prior on $\log_{10}(c_s^2)$; the results are shown in Fig.~\ref{fig:w_cs2_1D_2D}. As in this model $w$ has a much stronger effect on the CMB via the late-time ISW impacting the low-$\ell$ $C_\ell$, we can also present CMB-only results yielding
\beq
w = -0.91 ^{+0.34}_{-0.28}  \qquad (95\% \htwo \text{CL; CMB only} ) 
\eeq
for the background equation of state.  
 The previously discussed issue, with the perturbations being suppressed for $w \approx -1$, also persists in the EDE model, so we cannot make strong statements about $c_s^2$ from the CMB alone. Adding the cross-correlation data, the constraint on $w$ further tightens to 
\beq
w = -0.86 ^{+0.17}_{-0.16} \qquad (95\% \htwo \text{CL; CMB}+gg+Tg ) \, 
\eeq
without the $\varphi g$-CFs and
\beq
w = -0.80 \pm 0.18 \qquad (95\% \htwo \text{CL; CMB}+gg+Tg+\varphi g ) \, 
\eeq
with them. If we include the BAOs instead, we find
\beq
w = -0.92 ^{+0.12}_{-0.11}   \qquad (95\% \htwo \text{CL; CMB}+gg+Tg+\text{BAO} ) \, .
\eeq
It is important to note that these constraints are again different from those obtained in the previous models. In contrast to the $w$--$\log_{10}(\alpha)$ results for the entropy perturbation model, here we obtain significant lower limits on $\log_{10}(c_s^2)$. This is no surprise, as Figs.~\ref{fig:ClTT} to \ref{fig:Wth_Phi_LRG} have shown the stronger response of our observables in the anisotropic stress scenario.

 In the CMB-only case, the constraint is rather weak, but tightens to
\begin{align}
\log_{10}(c_s^2) & > -2.39 \qquad (95\% \htwo \text{CL; CMB}+gg+Tg ) \, ; \\
\log_{10}(c_s^2) & > -2.12 \qquad (95\% \htwo \text{CL; CMB}+gg+Tg+\varphi g ) \, ,
\end{align}
when the cross-correlation data set is included. When adding the BAOs, there is no constraint at the $2\sigma$-level; this is again due to $w \approx -1$.

\section{Conclusion} \label{sec:conclusion}
We have constrained generalised deviations from the standard model based on General Relativity (GR) with a cosmological constant using current structure formation data, including galaxy clustering, the ISW effect, and CMB lensing tomography.
In order to constrain gravity on the largest cosmological scales, it is crucial to combine probes of the expansion history, such as the BAO standard ruler, with tests of gravity at the perturbative level, since the gravitational field equations are necessarily modified if GR breaks down on large scales.
To keep our analysis as general as possible, we have assumed a recently proposed generic parameterisation of perturbations in DE/MG, which encompasses the departures from GR that would arise in the majority of physically viable models \cite{Battye2012, Battye2013a, Battye2013b}.

After illustrating the phenomenology of this parameterisation and the effect of the model parameters on the CMB and LSS observables,
we have used a combined ISW/LSS/CMB lensing data set to constrain linear DE/MG perturbations with an MCMC method. When combining the ISW/LSS/CMB lensing data with \Planck CMB measurements and baryonic acoustic oscillations, we first find a tight constraint on the background equation of state (assumed to be constant) of $w = 0.93 \pm 0.14$ (95\% CL; CMB+$gg$+$Tg$+BAO). This constraint however changes depending on the model we assume for the DE/MG perturbation evolution, and in some cases also depending on the choice of priors; this model dependency of $w$ is a first interesting result of our analysis.

In the case of the \emph{entropy perturbation model with only $w$ and $\alpha$}, we find a tendency towards  $w > -1$ and a low $\log_{10}(\alpha)$; e.g. $w=-0.83^{+0.19}_{-0.18}$ (95\%) and $\log_{10}(\alpha) < -1.45$ (68\% CL) from the CMB+$gg$+$Tg$ data. This would correspond to a clustering DE component; the significance of these results is however well below the $2\sigma$ level. If we open up the parameter space to account for the \emph{full entropy perturbation model}, this tendency disappears. We then find $w=-0.92^{+0.20}_{-0.16}$ (95\% CL) and $\log_{10}(\alpha) > -1.69$ (68\% CL) from the same data, in agreement with a cosmological constant. Furthermore we find upper limits of $\beta_1 < 1.45$ and $\beta_2 < 12.05$ (both 95\% CL). 
In the \emph{anisotropic stress scenario}, we obtain \mbox{$w=-0.86^{+0.17}_{-0.16}$} and also a significant upper limit of $\log_{10}(c_s^2) > -2.39$ (both 95\% CL, CMB+$gg$+$Tg$). Adding the BAO data enforces $w \approx -1$ and therefore significantly weakens the constraints on the DE/MG perturbation evolution parameters in all models.

All classes of DE/MG models that we tested reduce to a cosmological constant when $w \rightarrow -1$, so in that sense they are all viable extensions of the $\Lambda$CDM standard model. From a theoretical perspective, there is no compelling reason to prefer one over the other. Our work illustrates the importance of understanding and testing the DE/MG behaviour at the perturbative level, which will be the main goal for current and upcoming DE/MG missions.
 In case any deviation from a cosmological constant is found, the numerical value of $w$ and its interpretation will depend significantly on the assumed model of the dark sector perturbations.

The outlook for the future is mixed. On the CMB temperature side, the constraints will not improve significantly, as the sensitivity to DE/MG beyond the background equation of state only comes from the ISW on the largest angular scales, that are already limited by cosmic variance in \WMAP and \Planckc. CMB lensing data are however rapidly improving, as in this case most of the signal is on smaller angular scales. Current and future data e.g. from the South Pole Telescope will yield a significant increase in the signal-to-noise, that will result in improved DE/MG constraints at the perturbative level.
The outlook is also promising on the large-scale structure side: the next decade will bring an unprecedented wealth of high-precision data from surveys like DES, Euclid and LSST, so that we can expect the constraints on both the background expansion and the perturbation evolution to tighten considerably.

A more fundamental problem will however arise in case the next generation of LSS surveys will find the data to favour a constant $w \approx -1$ with much smaller uncertainty than today. Then the only statement we will be able to make about \nolinebreak[4] DE/MG is that it very closely mimics a cosmological constant, as in this limit the dark sector perturbations vanish. If so, we will be left with all the theoretical problems related to $\Lambda$, while lacking further tools to distinguish between different models.
 If however a future survey finds significant indications for a deviation from $w=-1$,  our model-independent scheme for constraining generalised DE/MG perturbations will be a valuable tool to select between the numerous existing models of cosmic acceleration.

\acknowledgments
We are grateful to Adam Moss and Jonathan Pearson for illuminating discussions, in particular on the implementation of the EDE model. We also thank Antony Lewis for helpful comments on modifying \textsc{Camb} and \textsc{CosmoMC}. Numerical computations were performed at the \emph{Rechenzentrum Garching} of the Max Planck Society.
BS gratefully acknowledges support from the \emph{German National Academic Foundation}, the \emph{Max Weber Programme of Bavaria} and an \emph{Isaac Newton Studentship}. JW, TG and BS also acknowledge support from the \emph{Transregional Collaborative Research Centre TRR 33 -- `The Dark Universe'}.

\bibliography{literaturepaper}
\bibliographystyle{JHEP}

\end{document}